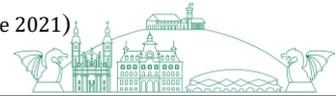

# DISTRIBUTION GRID ROBUST OPERATION UNDER FORECAST UNCERTAINTIES WITH FLEXIBILITY ESTIMATION FROM LOW VOLTAGE GRIDS USING A MONITORING AND CONTROL EQUIPMENT

Ankur Majumdar[1], Omid Alizadeh-Mousavi[1]

[1] – Depsys, Puidoux, Switzerland
ankur.majumdar@depsys.com

**Abstract** – Due to increased penetration of renewable resources in the distribution grid, the distribution system operator (DSO) faces increased challenges to maintain security and quality of supply. Since, a large proportion of renewables are intermittent generations, maintaining production and consumption balance of the electric system is at stake. The DSO needs to procure flexibilities not only from large resources, but also small-scale resources connected to the LV grid. Therefore, there is a need to estimate the aggregate flexibility at the secondary substation available from the LV grids. The flexibility estimation can only be achieved through monitoring of LV grid and the distributed energy resources (DERs) connected at the LV level. Moreover, the variability of intermittent resources affects the flexibility estimation methodology, and the DSO operation must be robust to the changes in generation or demand outputs. Furthermore, the up-to-date and accurate model of the LV grid is difficult to access. This paper demonstrates a methodology of estimating flexibility by sensitivity coefficients and thus, eliminating the need to have an up-to-date LV grid model. It further presents a robust optimisation based methodology for grid operation to address the impact of forecast uncertainties associated with the variable DER output. The methodologies illustrate the value of grid monitoring with a monitoring equipment. It has been validated on a real network of a Swiss DSO with several grid monitoring devices.

# 1 INTRODUCTION

The modern distribution grid is moving towards a decarbonised and decentralised system with increased number of renewable generations being connected at the medium voltage (MV) and low voltage (LV) distribution grids and increased electrification of demands such as, heating, cooling and transport. The distribution system operator (DSO), of today, is faced with a challenge to maintain security and quality of supply by procuring flexibilities from a multitude of resources such as, small-scale renewables, electric vehicles, battery energy storage systems, interruptible loads, thermostatically controlled loads connected to their grid [1].

To facilitate this, there is an increased requirement of digitalising the distribution grid with the integration of sensors equipped with information and communication technologies [2]. This has created opportunities for equipment with monitoring and control technologies. Furthermore, this will enable the DSOs to access and activate increased number of flexibilities and therefore, improve the security and quality of their grid operation.

The DSOs across the world are going through the process of smart meter roll-out [3]. However, the smart meters are mainly deployed for billing purposes. The privacy and access to smart meter data is a concern. In view of the general data protection regulation (GDPR), smart meter data from individual households are aggregated to the DSO after either anonymising or pseudonymising the data [4]. Besides, the 15-minute resolution data are communicated to the meter data management system only once or twice a day. Moreover, the DSOs will have to bear the high cost of hosting the higher resolution data in a meter data concentrator. All these barriers render the practicality of using the smart meter data for flexibility or near real-time applications difficult [5]. Therefore, there is a requirement for LV grid monitoring equipment which can not only address the data privacy issues but also, provide necessary platform to the DSO for access to high-resolution data for flexibility and near real-time applications.

Currently, the distribution grid is largely unmonitored at the LV level and limited monitoring at the MV level. Besides, the flexibilities available from the LV grid are limited without any high-resolution monitoring information. To maintain security and quality of supply and improve system operation, the grid operator must have information on aggregate flexibility estimation from the LV grid at the secondary substation.

This paper proposes a flexibility estimation methodology based on data of grid monitoring devices at the LV side of the secondary substation, the downstream LV street cabinets and DERs. The model-less sensitivity approach ensures that the LV grid remains secure with no voltage and congestion problems. The method is robust to changes in grid topology and operating points. This is particularly advantageous as there is little knowledge on up-to-date topology and parameters of the LV grid. The paper further proposes a robust optimisation approach capable to manage grid operation under forecast uncertainties with some predefined uncertainty levels and budget.

## 2 METHODOLOGY
### 2.1. CURRENT CHALLENGES IN THE GRID OPERATION

The modern distribution grid is generally not monitored at the LV level. At the MV level, monitoring is available only at the feeder-outs of the HV/MV transformer [6]. Due to the lack of monitoring information, the flexibility potential of LV grids and to a large portion of MV grid remains untapped. Furthermore, the current distribution system operation is based on load disaggregation in proportion to the ratings of MV/LV transformers [7]. This gives rise to sub-optimal operation, voltage violation and overloading of distribution grids. Furthermore, the flexibility from DERs is mainly realised by contractual flexibility or non-firm connection where, the DERs are contracted to disconnect whenever there is a security issue in the grid [8]. The flexibilities can also be realised by topological flexibility where, the topology of a distribution grid is changed through a soft-open point [9]. This, however, requires monitoring and communication technology and is thus, not popular with the DSOs. These severely limit the hosting capacity of flexible resources such as, renewables, energy storage, electric vehicles. The technical flexibilities available through optimal grid operation is largely not present in the current scenario as this requires significant grid visibility [1]. Nevertheless, with more and more decentralised resources being connected at the LV grid, the DSOs require more visibility and estimation of their flexibility potential at the LV grids.

### 2.2. PROPOSED APPROACH

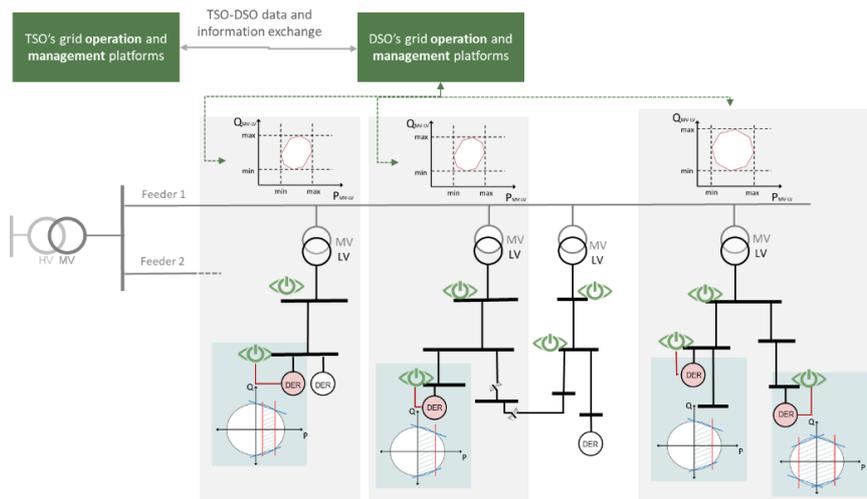

*Figure 1. Flexibility estimation and DSO operation and management platform*

The methodology of grid operation proposed in this paper is classified into ***two parts***. The ***first part*** is about estimating aggregate flexibilities at the secondary substation from the downstream LV grid, as shown in Figure 1. The flexibility estimation is a measure of how much deviation (flexibility) in terms of active and reactive power is allowed provided that the distribution LV grid is secure with no voltage violation or congestion. The methodology is realised by real-time and accurate monitoring data from grid monitoring equipment located at the secondary of the MV/LV transformer and the LV street cabinets. The flexibility estimation algorithm is hosted by the equipment located at the secondary of the MV/LV transformer. This can be further communicated to a central grid operational and management platform maintained by the DSO. Figure 1 also shows the capability curves of different DERs and the flexibility estimation capability curves at the MV/LV transformers. The *second part* is realising a robust approach of grid operation under forecast uncertainties. The approach is a robust optimisation-based power flow considering the worst-case scenarios of forecast uncertainties of DER outputs.

### 2.2.1. ROBUST LV OPF FOR FLEXIBILITY ESTIMATION FROM LV GRID

The aggregate flexibility estimation methodology (the *first part*)– is an optimisation framework maximising the flexibility capability area of active and reactive power flow through the MV/LV transformer in order to procure maximum possible flexibilities from the DERs. The impact of forecast errors on flexibility estimation is modelled by a robust optimisation framework considering the worst-case scenario of production or consumption uncertainties.

The formulation of the problem is an ***LV OPF*** problem as follows.

$$\max \quad \alpha \cdot \Delta p_{sl}^{LV} + \beta \cdot \Delta q_{sl}^{LV} \quad (1)$$
$$\text{s. to} \quad \Delta V_i^{LV} = K_{VP} * \Delta P_i^{LV} + K_{VQ} * \Delta Q_i^{LV} \quad (2)$$
$$\Delta I_{ij}^{LV} = K_{IP} * \Delta P_i^{LV} + K_{IQ} * \Delta Q_i^{LV} \quad (3)$$
$$V_i^{LV} = V_{i,0}^{LV} + \Delta V_i^{LV} \quad (4)$$
$$I_{ij}^{LV} = I_{ij,0}^{LV} + \Delta I_{ij}^{LV} \quad (5)$$
$$V^{min} \leq V_i^{LV} \leq V^{max} \quad (6)$$
$$\left|I_{ij}\right|^{LV} \leq I_{max} \quad (7)$$

Eq. (1) is the objective function maximising the active and reactive power flow through the MV/LV transformer subject to the linear power flow constraints (2)-(7). The linear power flow constraints are based on sensitivity coefficients $K_{VP}$, $K_{VQ}$ and $K_{IP}$, $K_{IQ}$ which denote the sensitivity of voltage and current respectively, calculated around the operating points of $V_{i,0}^{LV}$ for voltage and $I_{ij,0}^{LV}$ for branch current, with respect to change in active and reactive power injection or absorption. The sensitivity coefficients are hosted in the platform of grid monitoring devices located at the secondary of the MV/LV transformer together with those at the LV street cabinets. This ensures that the grid voltage and line flow always remain within their security limits [10]. The flexibility area (estimation) is determined by running the above optimal power flow several times based on the number of points required for achieving a complete P-Q flexibility capability area. The values of α and β is either 0 or ±1 depending on the direction of search in the P-Q flexibility capability area. The detailed algorithm has been reported in [11].

### 2.2.2. ROBUST MV OPF FOR DISTRIBUTION GRID OPERATION – ROBUST AGAINST FORECAST UNCERTAINTIES

Due to the lack of monitoring information at the LV grid, forecasting the intermittent resources for flexibility provision has been a challenge. Thus, having real-time and accurate grid monitoring data is particularly important for improved forecasts and improved flexibility provision.

Furthermore, it has been argued, in this scenario, that having interval forecasts with quantile distribution would make more sense rather than having point forecasts [12]. The system operation will also need to modify its approach of deterministic optimal operation to probabilistic optimal operation. This *second part* of the methodology presents one such approach where the operational platform of a DSO executes a robust optimisation-based *MV OPF* to take into account the intermittent nature of DER outputs by modelling them as interval forecasts. It is assumed here that the interval forecasts are available from accurate and granular grid monitoring data.

The formulation of the proposed MV OPF (as a second-order cone) is given below.

$$\min w_I \sum_{lines} r_{ij} l_{ij} + w_V \sum_{nodes} V_{dev} + w_{Ilim} \sum_{lines} l_{ij_{dev}} + w_p p_{sl}^{MV} + w_q q_{sl}^{MV} \quad (8)$$

$$\text{s.to} \quad \sum_{adj} P_{ij} + p_i = \bar{p}_i^g + \alpha . \max_{w \leq \Gamma} w . p_i^g - p_i^c \quad \forall \text{ node}, \forall \text{ adj lines} \tag{9}$$

$$\sum_{adj} Q_{ij} + q_i = \bar{q}_i^g + \alpha . \max_{w \leq \Gamma} w . q_i^g - q_i^c \quad \forall \text{ node}, \forall \text{ adj lines} \tag{10}$$

$$v_j = v_i - 2r_{ij}P_{ij} - 2x_{ij}Q_{ij} + (r_{ij}^2 + x_{ij}^2)l_{ij} \tag{11}$$

$$P_{ij}^2 + Q_{ij}^2 \leq v_i l_{ij} \tag{12}$$

$$p_i, q_i \in A_i^{MV/LV} \tag{13}$$

$$V_{dev} = 0 \text{ if } V^{min} \leq V_i \leq V^{max} \tag{14}$$

$$V_{dev} = |V_i|^2 - V_{lim}^2, \text{ if } V_i < V^{min} \text{ or } V_i > V^{max} \tag{15}$$

$$\left|I_{ij}\right|_{dev}^2 = 0 \text{ if } |I_{ij}| \leq I_{max} \tag{16}$$

$$\left|I_{ij}\right|_{dev}^2 = |I_{ij}|^2 - |I_{ij}|_{max}^2 \text{ if } |I_{ij}| > I_{max} \tag{17}$$

$$0 \leq \bar{p}_i^g + \alpha . \max_{w \leq \Gamma} w . p_i^g \leq p_{i,DER}^{max} \tag{18}$$

$$q_{i,DER}^{min} \leq \bar{q}_i^g + \alpha . \max_{w \leq \Gamma} w . q_i^g \leq q_{i,DER}^{max} \tag{19}$$

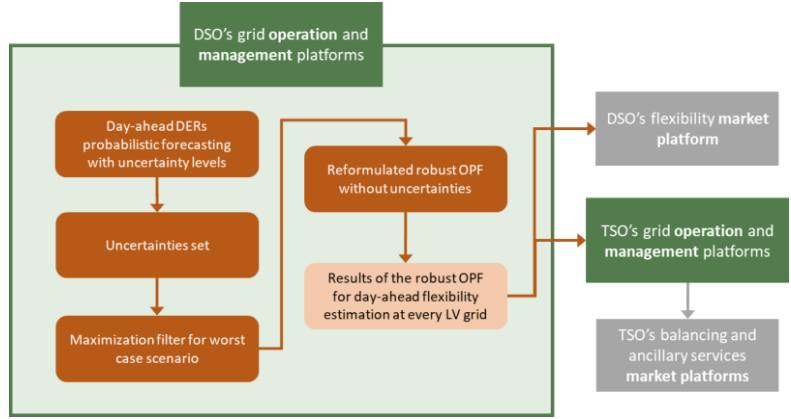

*Figure 2. Workflow for robust optimisation-based OPF with interval forecasts*

Eq. (8) is the objective function consisting of cost for losses, voltage and flow violations constrained by power flow constraints (9)-(12). The control setpoints of this MV OPF (*second part*) are constrained by flexibility capability area obtained in the *first part* (Eq. (13)). The uncertainties $w$ in relation to the DER output are bounded by a convex box set with uncertainty level α and uncertainty budget Γ (Eqs (18)-(19)) The robust optimisation considers the worst-case scenario of forecast variation from the predicted values. The robust optimisation, as addressed in the literature [13], a minimax problem, is usually solved by primal-dual optimisation. The minimax problem minimises the objective for the worst-case uncertainty. However, in this paper, an explicit maximisation filter method is used on the original robust counterpart of the problem to reformulate it to a counterpart with no uncertainties. The maximisation filter takes the worst case for the inequality constraints. The maximisation of the uncertainties for the inequality constraints turns it into a dual norm. Hence, the uncertainties are removed, and the robust optimisation is reformulated to a computable form. Figure 2 describes the procedure. The detailed explanation on the algorithm is given in [14].

## 3  RESULTS

The proposed methodology is tested and validated in a real network of a Swiss DSO. The test setup considers that there are LV grid monitoring and control equipment with predefined interval forecasts associated with DER (PV) output.

The case study considers two scenarios – ***today scenario*** with current loading condition and ***future scenario*** with increased uncontrolled loading of 100kW at each LV grid due to electrification of heating, loading and transport. The test setup is simulated by running the sensitivity coefficients-based LV OPF (***first part***) to estimate the aggregate PQ flexibility at each MV/LV transformer and then executing a robust MV OPF (***second part***) considering the worst-case scenario for DER forecasts with uncertainty level α=0.5.

Figure 3 and Figure 4 show the test MV network and a LV grid of the Swiss DSO, respectively. The GridEye monitoring devices installed at the seocndary of the MV/LV transformers and the LV cabinets are also shown in the figures. The grid monitoring devices have accurate 10-min data on voltage, current, active and reactive powers.

The LV grids of the test network have a total of 200kWp of PVs installed. It is assumed that the installed

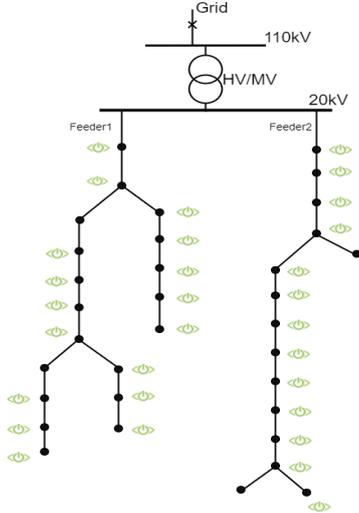

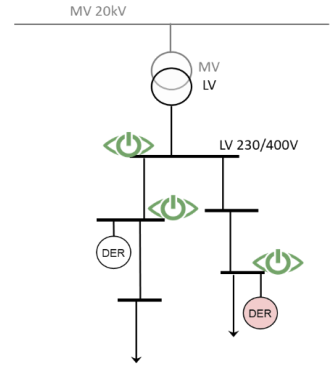

*Figure 4. A representative LV grid with DERs and monitoring devices.*

*Figure 3. MV grid topology and monitoring devices.*

PVs have flexibility potential of up to 10% curtailment of active power with pf control. Figure 5 illustrates the aggregate flexibilities of a particular LV grid with the worst-case uncertainties ($\alpha=\pm0.5$) compared to the expected/predicted output ($\alpha=0$) for *today scenario*. Figure 6 further shows the comparison of losses in kWh and grid violation (voltage+flow) costs in CHF for three values of $\alpha$.

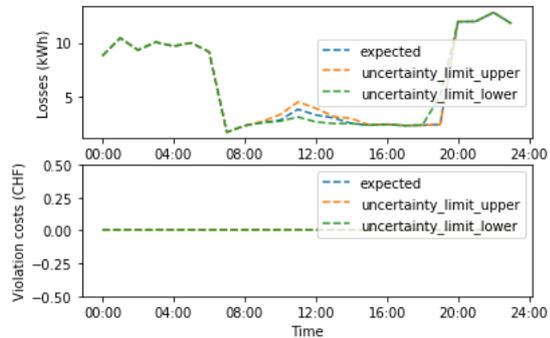

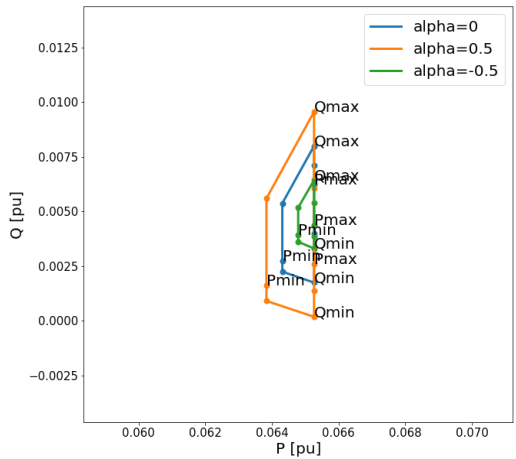

*Figure 6. Losses and security violation costs for the predicted and the worst-case uncertainties for today scenario*

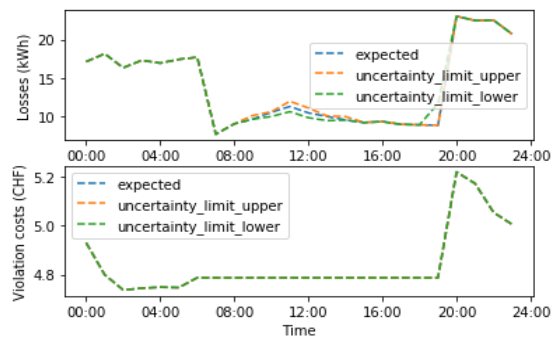

*Figure 5 PQ flexibility area for the worst-case uncertainties and the point forecast for today scenario*

*Figure 7. Losses and security violation costs for the predicted and the worst-case uncertainties for future scenario*

Similarly, for the *future loading scenario*, Figure 7 shows the losses in kWh and grid violation costs in CHF (calculated with the violations from their limits being multiplied by 100 CHF) for the three values of α. It is observed that there is 1.65% and 0.43% increase in losses for the worst case upper and lower limits of forecast variations respectively, for *today scenario* and 0.71% and 0.11% increase in losses for the worst case upper and lower limits forecast variations respectively, in *future scenario* compared to the expected forecasts. The increase in total losses for the *future scenario* with expected forecasts compared to the *today scenario* is 182 kWh. It is observed that there are no violation costs reported in the first scenario. There is a security violation costs of only about 116.176 CHF for the *future scenario* with uncontrolled load increase. This shows that secure operation of the grid can be significantly improved with this approach for future loading scenarios.

# 4 CONCLUSIONS

The combined methodology of aggregate flexibility estimation from the LV grid and the robust MV grid operation under uncertainties is beneficial for many reasons. Firstly, this will enable the DSOs to have a measure of aggregate flexibilities available from the LV grid. Secondly, it will ensure the secure MV grid operation under characterised forecast uncertainties from DERs. Thirdly, this will facilitate demand response programs through electric vehicles, thermostatically controlled loads, energy storages, and installed capacity of renewables in the grid. Fourthly, the proposed methodology employs a modelled approach only for the MV grid and does not require an updated LV grid model. The simulation and results indicate that LV monitoring and control plays an important role not only in improved interval forecast parameterisation but also enables the secure operation of distribution grid under the worst-case uncertainties. As a future step, the flexibility estimation algorithm will be carried out for the primary substation level to realise the aggregate flexibilities provided to TSO ancillary and balancing services.

# REFERENCES

[1] MILLS G., & MACGILL I.: (2018). Assessing electric vehicle storage, flexibility, and distributed energy resource potential. Journal of Energy Storage, 17, 357-366.
[2] EUROPEAN COMMISSION.: ICT for a Low Carbon Economy Smart Electricity Distribution Networks, https://bit.ly/3qJRQwT, accessed Feb 2021
[3] EUROPEAN COMMISSION.: Smart Metering deployment in the European Union, http://bit.ly/2MbDfeL, accessed Feb 2021
[4] SSEN.: Smart Meter Data Privacy Plan https://bit.ly/3s6QMDE accessed Feb 2021
[5] SOVACOOL B. K., KIVIMAA P., HIELSCHER S., & JENKINS K.: (2017). Vulnerability and resistance in the United Kingdom's smart meter transition. Energy Policy, 109, 767-781.
[6] NAYEL M., MORAD M., & MOHAMED W.: (2019, November). Monitoring of Electric Distribution Grids: Existing, Priority, and Objective. In 2019 IEEE Sustainable Power and Energy Conference (iSPEC) (pp. 2182-2186). IEEE.
[7] CARMONA C., ROMERO-RAMOS E., RIQUELME J., & GOMEZ-EXPOSITO A. (2010, October). Distribution transformer load allocation from substation measurements and load patterns. In 2010 IEEE PES Innovative Smart Grid Technologies Conference Europe (ISGT Europe) (pp. 1-8). IEEE.
[8] NIE Non-Firm Access Connections for Distribution Connected Distributed Generators, https://bit.ly/2NpgdBu, accessed Fen 2021
[9] LONG C., WU J., THOMAS L., & JENKINS N.: (2016). Optimal operation of soft open points in medium voltage electrical distribution networks with distributed generation. Applied Energy, 184, 427-437.
[10] CHRISTAKOU K., LEBOUDEC J. Y., PAOLONE M., & TOMOZEI D. C. (2013). Efficient computation of sensitivity coefficients of node voltages and line currents in unbalanced radial electrical distribution networks. IEEE Transactions on Smart Grid, 4(2), 741-750.
[11] SILVA J., SUMAILI J., BESSA R. J., SECA L., MATOS M. A., MIRANDA V., ... & SEBASTIAN-VIANA M. (2018). Estimating the active and reactive power flexibility area at the TSO-DSO interface. IEEE Transactions on Power Systems, 33(5), 4741-4750.
[12] PINSON P., KARINIOTAKIS G., NIELSEN H. A., NIELSEN T. S., & MADSEN H. (2006, February). Properties of quantile and interval forecasts of wind generation and their evaluation. In Proceedings of the European Wind Energy Conference & Exhibition, Athens (pp. 1-10).
[13] THIELE A., TERRY T., & EPELMAN M.: (2009). Robust linear optimization with recourse. Rapport technique, 4-37.
[14] BAI X., & QIAO W.: (2015). Robust optimization for bidirectional dispatch coordination of large-scale V2G. IEEE Transactions on Smart Grid, 6(4), 1944-1954.